\documentclass[12pt]{iopart}
\usepackage{iopams}  

\usepackage{graphicx}
\usepackage{subfigure}
\usepackage[]{url}
\usepackage{verbatim}

\newcommand{\degree}{\ensuremath{^\circ}}

\begin{document}

\title[Test of the gravitational redshift, application to Galileo satellites 5\&6]{Test of the gravitational redshift with stable clocks in eccentric orbits: application to Galileo satellites 5 and 6}

\author{ P. DELVA$^1$, A. HEES$^2$, S. BERTONE$^3$, E. RICHARD$^1$ AND P. WOLF$^1$ }
\address{$^1$ SYRTE, Observatoire de Paris, PSL Research University, CNRS, Sorbonne Universit\'es, UPMC Univ. Paris 06, LNE, 61 avenue de l'Observatoire, 75014 Paris, France\\
$^2$ Department of Mathematics, Rhodes University, 6140 Grahamstown, South Africa\\
$^3$ Astronomical Institute, University of Bern, Sidlerstrasse 5 CH-3012 Bern, Switzerland}


\begin{abstract}
The Einstein Equivalence Principle (EEP) is one of the foundations of the theory of General Relativity and several alternative theories of gravitation predict violations of the EEP. Experimental constraints on this fundamental principle of nature are therefore of paramount importance. The EEP can be split in three sub-principles: the Universality of Free Fall (UFF), the Local Lorentz Invariance (LLI) and the Local Position Invariance (LPI). In this paper we propose to use stable clocks in eccentric orbits to perform a test of the gravitational redshift, a consequence of the LPI. The best test to date was performed with the Gravity Probe A (GP-A) experiment in 1976 with an uncertainty of $1.4\times10^{-4}$. Our proposal considers the opportunity of using Galileo satellites 5 and 6 to improve on the GP-A test uncertainty. We show that considering realistic noise and systematic effects, and thanks to a highly eccentric orbit, it is possible to improve on the GP-A limit to an uncertainty around $(3-4)\times 10^{-5}$ after one year of integration of Galileo 5 and 6 data.
\end{abstract}

%
\vspace{2pc}

%
%
%
\section{Introduction}
The classical theory of General Relativity (GR) is the current paradigm to describe the gravitational interaction. Since its creation in 1915, GR has been confirmed by experimental observations. Although very successful so far, it is nowadays commonly admitted that GR is not the ultimate theory of gravitation. Attempts to develop a quantum theory of gravitation or to unify gravitation with the others fundamental interactions lead to deviations from GR. Moreover, observations requiring the introduction of Dark Matter and Dark Energy are sometimes interpreted as a hint that gravitation presents some deviations from GR at large scales.

GR is built upon two fundamental principles. The first principle is the Einstein Equivalence Principle (EEP) which gives to gravitation a geometric nature. More precisely, the EEP implies that gravitation can be identified with space-time curvature which is mathematically described by a space-time metric $g_{\mu\nu}$. If the EEP postulates the existence of a metric, the second principle of GR specifies the form of this metric. In GR, the metric tensor is determined by solving the Einstein field equations which can be derived from the Einstein-Hilbert action. 

From a phenomenological point of view, three aspects of the EEP can be tested~\cite{Will1993}: (i) the Universality of Free Fall (UFF), (ii) the Local  Lorentz Invariance (LLI) and (iii) the Local Position Invariance (LPI). The UFF has been tested experimentally at the very impressive level of $10^{-13}$ with torsion balances (see e.g. \cite{schlamminger:2008zr,adelberger:2009zr}) and with Lunar Laser Ranging measurements \cite{williams:2009ys}. One way to test the LLI is to search for anisotropies in the speed of electromagnetic interactions, usually parametrized with the $c^2$-formalism~\cite{Will1993}. It is interesting to mention that GPS data have provided a good test of the LLI \cite{wolf:1997fk}, even if a better constraint is now given by Ives-Stilwell type experiments~\cite{Botermann2014}. The last part of the EEP, the LPI is tested by constraining space-time variations of the constants of Nature (see e.g. \cite{rosenband:2008fk,uzan:2011vn,guena:2012ys,hees:2014uq}) or by redshift tests \cite{vessot79,vessot80,vessot89}.

In addition to the EEP tests, the second principle of GR (the form of the metric) is thoroughly tested by different observations in the Solar System: deflection of light \cite{lambert:2009bh}, planetary ephemerides \cite{verma:2014jk,fienga14,pitjeva:2014fj}, radioscience tracking data of spacecraft \cite{bertotti:2003uq,hees:2014jk,hees:2012zr}, etc (for a larger review, see \cite{Will1993,Will2014}). 

Tests of the EEP are of prime importance since violations thereof are predicted by attempts to develop a quantum theory of gravity or a theory unifying all fundamental interactions and in several models of Dark Energy (see e.g. \cite{damour:2012zr} and references therein for some examples). The level of the EEP violation depends highly on the theoretical model considered~(see for example~\cite{damour:2010ly}) while the different types of EEP tests (universality of free fall, redshift test, constancy of the constants of Nature) are sensitive to different fundamental parameters~\cite{Damour1999a} and are very complementary (see also the discussion in section V.A. of~\cite{altschul:2015sh}).

As mentioned above, a gravitational redshift experiment tests the LPI~\cite{Will1993}.  The most precise test of the gravitational redshift to date has been realized with the Vessot-Levine rocket experiment in 1976, also named the Gravity Probe A (GP-A) experiment~\cite{vessot79,vessot80,vessot89}. The frequency differences between a space-borne hydrogen maser clock and ground hydrogen masers were measured thanks to a continuous two-way microwave link.  The gravitational redshift was verified to $1.4\times 10^{-4}$ accuracy~\cite{vessot89}.  The future Atomic Clock Ensemble in Space (ACES) experiment, an ESA/CNES mission, planned to fly on the ISS in 2017, will test the gravitational redshift to around $2-3\times 10^{-6}$ accuracy~\cite{ACES2009}. Furthermore, other projects like STE-QUEST \cite{altschul:2015sh} propose to test the gravitational redshift at the level of $10^{-7}$, and observations with the RadioAstron telescope may reach an accuracy of the order of $10^{-5}$~\cite{litvinov:2015lh}. Finally, it has been previously suggested in~\cite{Svehla2010} to use Galileo satellites for such a test.

In this paper, we will show how clocks on-board GNSS satellites can be used to perform an improved test of the gravitational redshift if they are placed on an elliptical orbit around Earth.   As an application, we will consider the case of the satellites Galileo 5 and 6. These satellites were launched on August, 30th 2014 and because of a technical problem, the launcher brought them on a wrong, elliptic orbit. An elliptic orbit induces a periodic modulation of the gravitational redshift \cite{ashby:2003wb} while the good stability of recent GNSS clocks allows to test this periodic modulation to a very good level of accuracy. The Galileo 5 and 6 satellites, with their large eccentricity and on-board H-maser clocks, are hence perfect candidates to perform this test. Contrary to the GP-A experiment, it is possible to integrate the signal on a long duration, therefore improving the statistics. In this paper, we assess the sensitivity of the gravitational redshift test that can be performed with Galileo satellites~5 and~6.

In Sec.~\ref{sec:expdata}, we describe the experimental data needed in order to perform the test with Galileo satellites. Then, we introduce in Sec.~\ref{sec:simu} the simulation of the satellites orbits, clocks and signal. In Sec.~\ref{sec:stat} we estimate the test sensitivity with two different statistical methods to estimate the averaging of random noise in the signal. Finally, in Sec.~\ref{sec:syst}, we present a detailed study of the different systematics that need to be modeled in our analysis and we show how to decorrelate them from the gravitational redshift signature.


\section{Experimental data}\label{sec:expdata}

The experiment proposed here requires an accurate knowledge of the frequency of the satellite clock as it orbits the Earth.
These data are made available by several analysis centers (ACs) of the International GNSS Service (IGS,~\cite{2009JGeod..83..191D}) in the framework of the Multi-GNSS-EXperiment (MGEX,~\cite{Montenbruck2013GPSWorld}). 
MGEX applies the standard IGS processing chain to new GNSS (Galileo, Beidu) and regional services (e.g., QZSS for Japan).
All active Galileo satellites orbits, clock corrections, and inter-system biases are then available for public use through the MGEX product directory at the IGS data center CDDIS, which is updated on a regular basis.

Since each AC is following a different strategy, as an example we shall focus on the processing performed at the CODE (Center for Orbit Determination in Europe) analysis center~\cite{Dach2013}. 
In this particular case, the whole solution is based on a two steps process. First, satellite orbits are determined using a double difference network solution. This technique allows to eliminate or reduce several biases by combining observations among a pair of receivers and a pair of satellites. All clock errors are eliminated in the combination while the remaining parameters (including ephemerides, atmospheric and antenna parameters, etc...) are estimated for all stations and satellites involved. The latter are then used as a framework for the clock solution, which is based on a zero-difference processing. For more general information about the IGS and its products and data centers, the reader can refer to~\cite{2009JGeod..83..191D}.

For the test proposed in this paper,  the orbit solution of Galileo 5 and 6 satellites would be used to calculate the behaviour of the on-board clocks and the gravitational redshift as predicted by GR~\cite{Will2014}. The latter would then be compared to the clock solution from the IGS processing to recover any violation of the LPI, as detailed in Sec.~\ref{sec:stat}.  Several conditions need to be satisfied by the data to assure the success of the experiment. The availability of continuous orbit and clock parameters is paramount and is assured by the IGS guidelines for the clock and orbit products. Orbits and station coordinates should be unaffected by an eventual violation of the gravitational redshift. This is assured as a result of the processing outlined above, since both are derived by a double-difference solution which eliminates all clock parameters.
Finally, the reference clock for the solution is always chosen among ground station H-maser clocks~\cite{2009JGeod..83.1083B}, so that it is unaffected by an eventual LPI violation.

\section{Simulation: orbits, clocks and signal}\label{sec:simu}
\begin{figure}[t]
   \begin{minipage}[c]{.46\linewidth}
	   \includegraphics[width=1.0\linewidth]{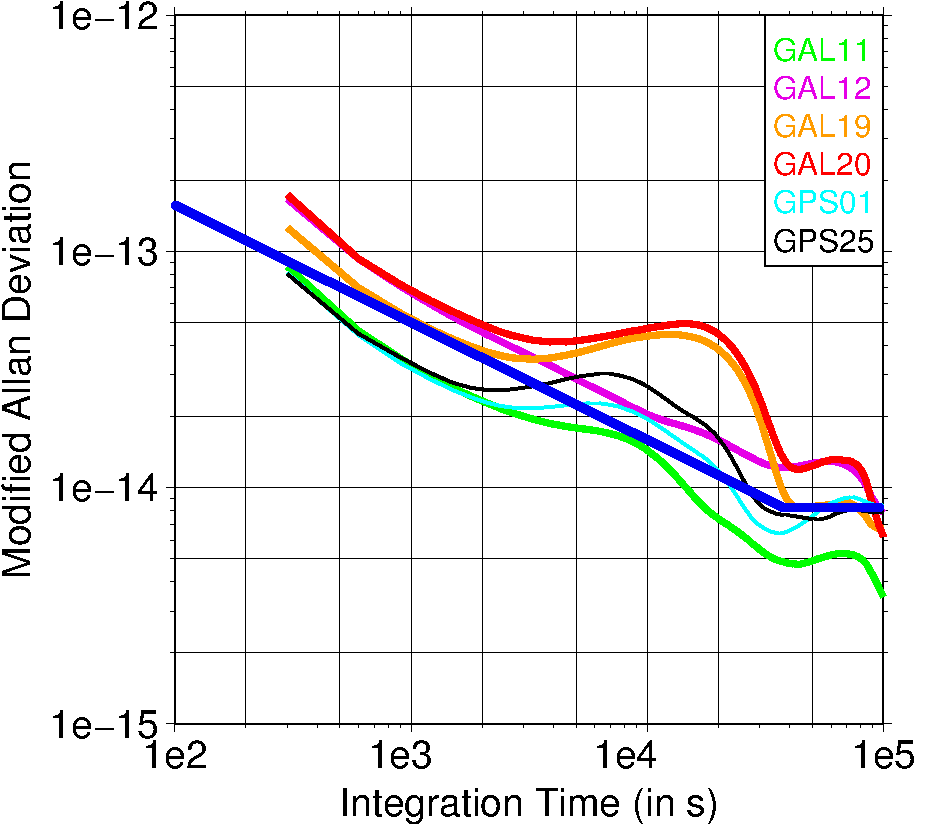}
   \end{minipage} \hfill
   \begin{minipage}[c]{.46\linewidth}
      \includegraphics[width=1.0\linewidth]{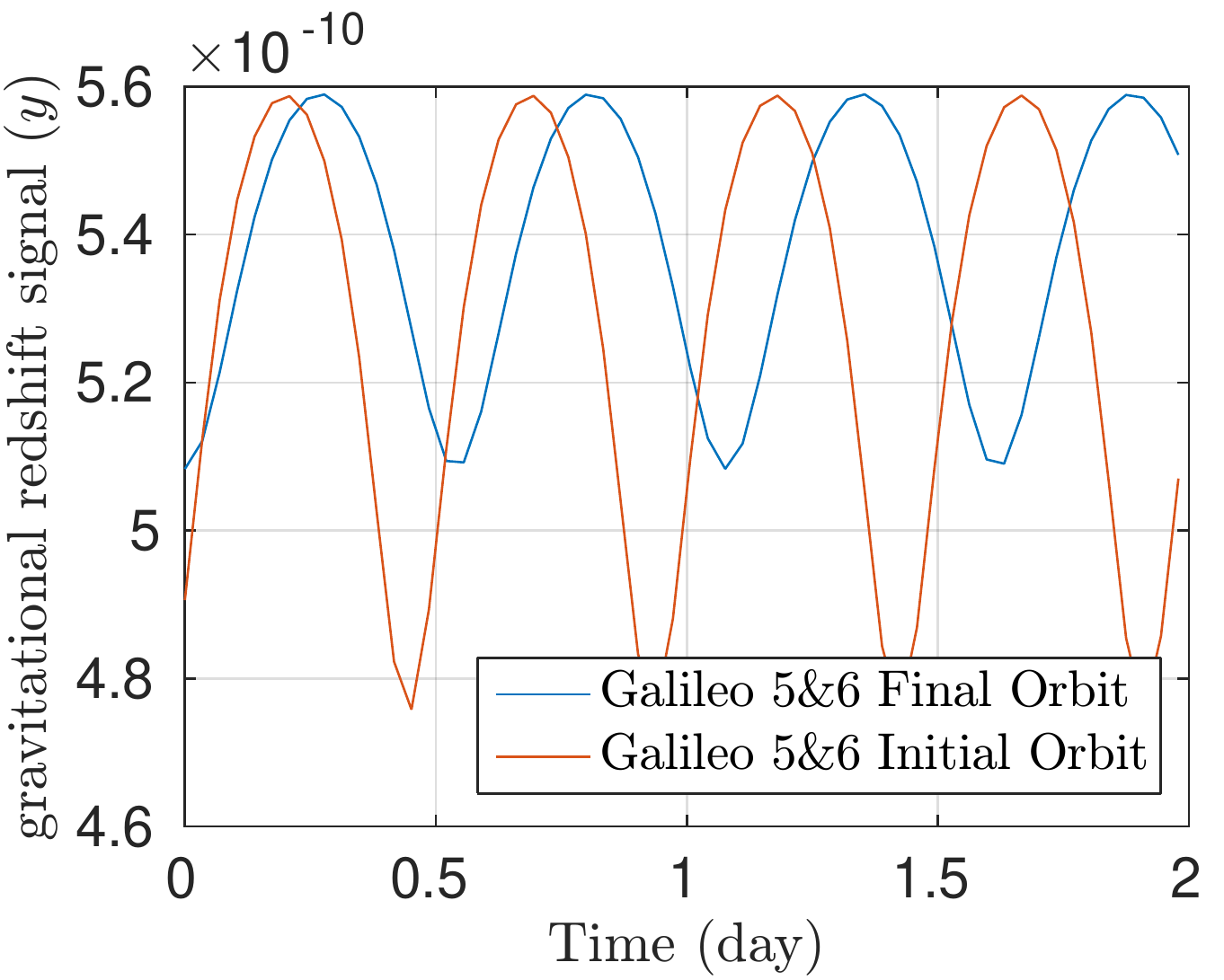}
   \end{minipage}
   \caption{\label{fig:clk} (left) MDEV for Galileo In-Orbit Validation (IOV) (GAL11-12-19-20) and for GPS Block II-F (PRN GPS01-25) from~\protect\cite{prange2014}. Superimposed in blue is the MDEV of the simulated clock noise; (right) Gravitational redshift signal with $\alpha = 0$.}
\end{figure}

\begin{table}[b]
\begin{center}
	\begin{tabular}{c|c|c|c}
Name & $e$ & $a$ (km) & $i$ ($\degree$) \\
\hline
Galileo 5\&6 Final Orbit & 0.1561 & 27977 & 49.7212 \\
Galileo 5\&6 Initial Orbit & 0.2330 & 26192 & 49.7740 
\end{tabular}
\caption{\label{tab:sat} Orbital parameters of satellites Galileo 5 and 6~\cite{ESA_pers}. Their orbits were circularized so that they can be used for positioning. Both orbits are now very similar.}
\end{center}
\end{table}

Orbital parameters of Galileo satellites 5 and 6 used for the simulation of the present work are summarized in table~\ref{tab:sat}. Their orbits have been recently circularized, from eccentricity $e=0.2330$ to $e=0.1561$. However, we will consider both the initial and the final orbits, in order to compare the sensitivity of the redshift test to different eccentricities. Orbits are calculated by solving the Kepler equation for a duration of two years, thus ensuring the numerical stability of the solution with time.

We use the clock solutions provided by CODE to simulate a realistic random noise for Galileo satellite clocks. Fig.~\ref{fig:clk} (left) shows the Modified Allan Deviations (MDEVs) for several GPS and Galileo clocks~\cite{prange2014} around Day Of Year (DOY) 100/2013, as well as the MDEV of the simulated clock noise (in blue). For an integration time of 1000~s, the MDEVs are between $3\times 10^{-14}$ and $7\times 10^{-14}$, and the MDEV of the simulated clock noise is around $5\times 10^{-14}$. The MDEVs decrease like $\tau^{-1/2}$, which is a characteristic of white frequency noise. There is a bump at about 21000~s due to systematic effects, which will be analysed in Sec.~\ref{sec:syst}. Moreover, we add a flicker noise to the simulated clock noise at $8\times 10^{-15}$ in the MDEV (flat part of the blue curve on Fig.~\ref{fig:clk}, left). Although Fig.~\ref{fig:clk}~(left) shows no sign of an increase of noise at large averaging times (indicative of random walk frequency noise or frequency drift), on ground measurements of Galileo H-maser clocks have shown such behaviour~\cite{rocha:2012rz}. The latter is anyway not exceeding $5\times10^{-15}$ in MDEV for measurements up to $10^5$~s averaging time. At the frequency of our expected periodic signal ($2.1\times10^{-5}$~Hz), the corresponding noise is about an order of magnitude below our noise model, and we will therefore neglect it in our analysis.

The ideal signal $y$ is the gravitational part of the relative frequency difference between a ground clock $g$ and the satellite clock $s$:
\begin{equation}
\label{eq:sig}
y = \frac{GM}{c^2} \left( \frac{1}{r_g} - \frac{1}{r_s} \right) \ ,
\end{equation}
where $GM$ is the standard gravitational parameter of Earth, $c$ is the speed of light in vacuum, and $r_g$ and $r_s$ are respectively the norm of the position vectors of the ground clock and of the on-board clock. We neglect the noise coming from the ground clock, and we are interested only in the variable part of $y$. Therefore one can remove the first term in~(\ref{eq:sig}), which is constant.

We use a simple phenomenological model for a possible violation of the gravitational redshift, characterised by a parameter $\alpha$ which is zero in general relativity~\cite{Will1993,Will2014}:
\begin{equation}
\label{eq:viol}
\tilde y (\alpha) = - (1+\alpha) \frac{GM}{c^2 r_s} \ .
\end{equation}
The GP-A limit corresponds~\cite{vessot89} to $\sigma_\alpha = 1.4 \times 10^{-4}$, where $\sigma_\alpha$ is the uncertainty on the estimation of the parameter $\alpha$.

The simulated signal $y$ is shown in Fig.~\ref{fig:clk}~(right). Its amplitude is around twice larger for Galileo satellites~5\&6 initial orbits than for Galileo~5\&6 final orbits. The asymmetry between the upper part and the bottom part of the curves is a characteristic of the elliptical orbit. 

\section{Estimation of the statistical sensitivity of the test} \label{sec:stat}

\begin{figure}[t] 
\begin{center}
\includegraphics[width=0.8\linewidth]{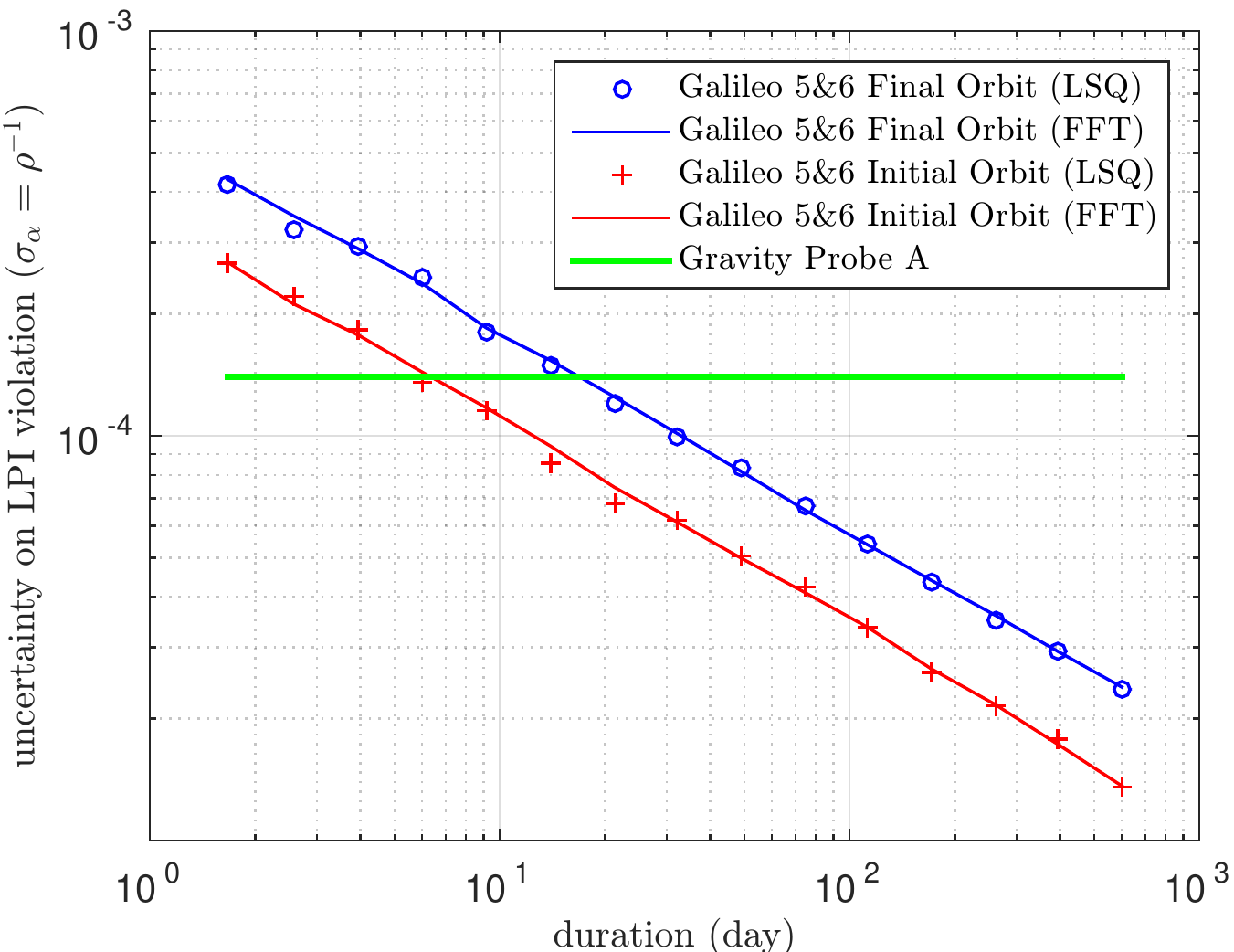}
\caption{Statistical sensitivity of the gravitational redshift test with respect to the duration of the experiment. We show the results for the two methods FFT and LSQ, and for the initial and final orbits of Galileo satellites~5\&6. In green is the accuracy of the Gravity Probe A experiment for reference.
\label{fig:sensit}}
\end{center}
\end{figure}

We study the sensitivity of the gravitational redshift test performed with Galileo~5 and~6 through two very different methods: the Fast Fourier Transform (FFT) method and the Linear Least-Square (LSQ) method.

The sensitivity of the gravitational redshift test can be calculated as the inverse of the Signal-to-Noise Ratio (SNR) $\rho$. The SNR can be maximized with matched filtering, i.e. by using a filter optimized for the searched signal and the considered noise. Then the SNR is given by~\cite{gardner1990}:
\begin{equation}
\label{eq:sn}
\rho^2 = \int_{-\infty}^{+\infty}
\frac{|\mathcal F[y] (f) |^2}{S_\epsilon(f)} \textrm{d} f \ ,
\end{equation}
where $y$ and $\epsilon$ are the gravitational redshift signal and the clock noise, respectively, $\mathcal F[y]$ is the Fourier transform of $y$, and $S_\epsilon$ is the Power Spectral Density (PSD) of the clock noise. We implemented a numerical version of this method with the following steps: first we generate 100 different sequences of clock noise; we calculate the PSD of each sequence using a fast Fourier transform algorithm and average the 100 different PSD; finally we compute the Signal-to-Noise Ratio (SNR)~(\ref{eq:sn}) with the mean PSD of the clock noise, by integrating in the frequency domain $[-F_s/2,F_s/2]$, where $F_s$ is the sampling rate of the signal. This is done for several different durations of the experiment.

The second method implemented consists in using a linear least-square (LSQ) fit combined with a Monte-Carlo method, where we find the minimum of the merit function
\begin{equation} \label{eq:chisq}
\chi^2 = \sum_{i=1}^N \left[ (y(t_i) + \epsilon_i) - ( \tilde y(\alpha; t_i) + A ) \, \right]^2 ,
\end{equation}
with respect to $\alpha$ and $A$. In Eq.~(\ref{eq:chisq}) $N$ is the number of simulated observables,  $A$ is a constant to be estimated and $(\epsilon_i)_{i=1..N}$ is one sequence of simulated clock noise, while $(y(t_i) + \epsilon_i)$ corresponds to the simulated observable at time $t_i$. The constant $A$ is introduced to remove a constant frequency bias which cannot be measured accurately.
Instead of fitting $A$, it is also possible to remove the mean of $\tilde y(\alpha)$ from the model, so that the fitted model has a zero mean, like the noise $\epsilon$.
We implemented a numerical version of this method, with the following steps: we generate 100 different sequences of clock noise (we take the same ones as for the FFT method); we estimate $\alpha$ for each clock noise sequence with the linear LSQ method, for different durations of the experiment; finally, for each duration, we calculate the mean of the 100 obtained values of $\alpha$ and their standard deviation~$\sigma_\alpha$. The standard deviation corresponds approximately to 68\% of the obtained values around the mean.


Using~\cite{arun2005}, it can be shown that for high SNR and in the case of a one parameter determination, the inverse of the SNR found with matched filtering method is equal to the standard deviation of $\alpha$ found with the LSQ method, i.e. $\sigma_\alpha = \rho^{-1}$.

Fig.~\ref{fig:sensit} shows the statistical sensitivity of the gravitational redshift test, $\sigma_\alpha$, with respect to the duration of the experiment for the two different methods presented above and for the final and initial orbits of Galileo satellites 5~\&~6. Also, we verify that the two methods give the same estimate, thus giving us strong confidence in  the reliability of our analysis.
This result is very promising, as it can be seen that even with the less eccentric orbit, the GP-A limit could be attained in less than a month. However, this result has to be tempered with a study of systematic effects.

\section{Systematic effects} 
\label{sec:syst}

All systematic effects (i.e. errors that are due to unmodeled deterministic effects) that mimic the gravitational redshift signal may induce a bias in the estimation of $\alpha$, i.e. a ``fake'' violation of the gravitational redshift. This will ultimately limit the useful duration of noise averaging presented in Fig.~\ref{fig:sensit}. The bias on the estimated value of $\alpha$ resulting from unmodeled systematic effects will depend on their magnitude, frequency and phase.
We classify systematic effects in four classes: (i) effects acting on the frequency of the reference ground clock; (ii) effects on the links, e.g., mismodeling of ionospheric or tropospheric delays, variations of receiver/antenna delays, multipath effects, etc... ; (iii) effects acting directly on the frequency of the space clock, e.g., temperature and/or magnetic field variations on board the Galileo satellites ; (iv) orbit modelling errors. The latter directly translate into variations of the measured clock frequencies because the clock solutions are obtained from the one-way signals (so called ``zero differences"), hence an orbit error of $\Delta x$ along the line of sight leads directly to an error in the clock phase estimation of $\simeq\Delta x/c$.  

Effects on the ground reference clock are likely to be negligible with respect to the others, and are furthermore likely to be only weakly correlated with the signal we are searching for shown in Fig. \ref{fig:clk}~(right). They will therefore be neglected in this analysis.

Effects on the link could be significant for certain stations (e.g. under unfavourable multipath environment or atmospheric conditions). However they are expected to vary with elevation and azimuth, and local meteorological conditions, none of which are correlated with the expected signal. Note also that the main signal frequency is $2.1\times 10^{-5}$~Hz whereas the 2nd harmonic of any diurnal effect (temperature, humidity etc...) is $2.3\times 10^{-5}$~Hz, which indicates that such effect will decorrelate from the signal with as little as 6 days of data. Finally, some more significant averaging can be expected given the large number of observing stations.

The environmental sensitivity of the Galileo passive hydrogen masers (PHM) has been characterized on ground (see e.g.,~\cite{rocha:2012rz}). Temperature sensitivity is $<2\times 10^{-14}/$K and magnetic field sensitivity $<3\times 10^{-13}/$G. Both Galileo 5 and 6 are equipped with on-board temperature sensors, but no in-situ temperature measurements are publicly available. From Fig.~\ref{fig:clk}~(right) we see that the expected peak to peak variation of the redshift signal is about $5\times 10^{-11}$. Given the expected statistical sensitivity of our test (cf. Fig.~\ref{fig:sensit}) we want any systematic effect to remain $2\times 10^{-5}$ below that level, i.e. $\leq 10^{-15}$. 
This implies that temperature variations at the clock that are correlated with the signal need to be $\leq 0.05$~K peak to peak. Whether this is achieved in orbit remains to be seen once temperature measurements are made available. As an example, \cite{Montenbruck2012} mention that for the GPS block IIF ``the rubidium clocks of the Block IIF satellites are mounted on a thermally isolated
sub-panel (along with the frequency distribution unit) for which a peak-to-peak temperature variation of less than 0.5 K has been observed in SVN62 on-board measurements. In addition, the new clock benefits from a baseplate temperature controller (BTC), which further reduces the sensitivity to external temperature variations by up to a factor of 50.'' This indicates that peak to peak variations $\leq 0.05$~K are realistic. Additionally, significant decorrelation between temperature effects and our signal can be expected, as the former depend on Sun exposure of the satellite and thus have annual spectral components, which are absent in the signal.

\begin{figure}[t] 
\begin{center}
\includegraphics[width=0.8\linewidth]{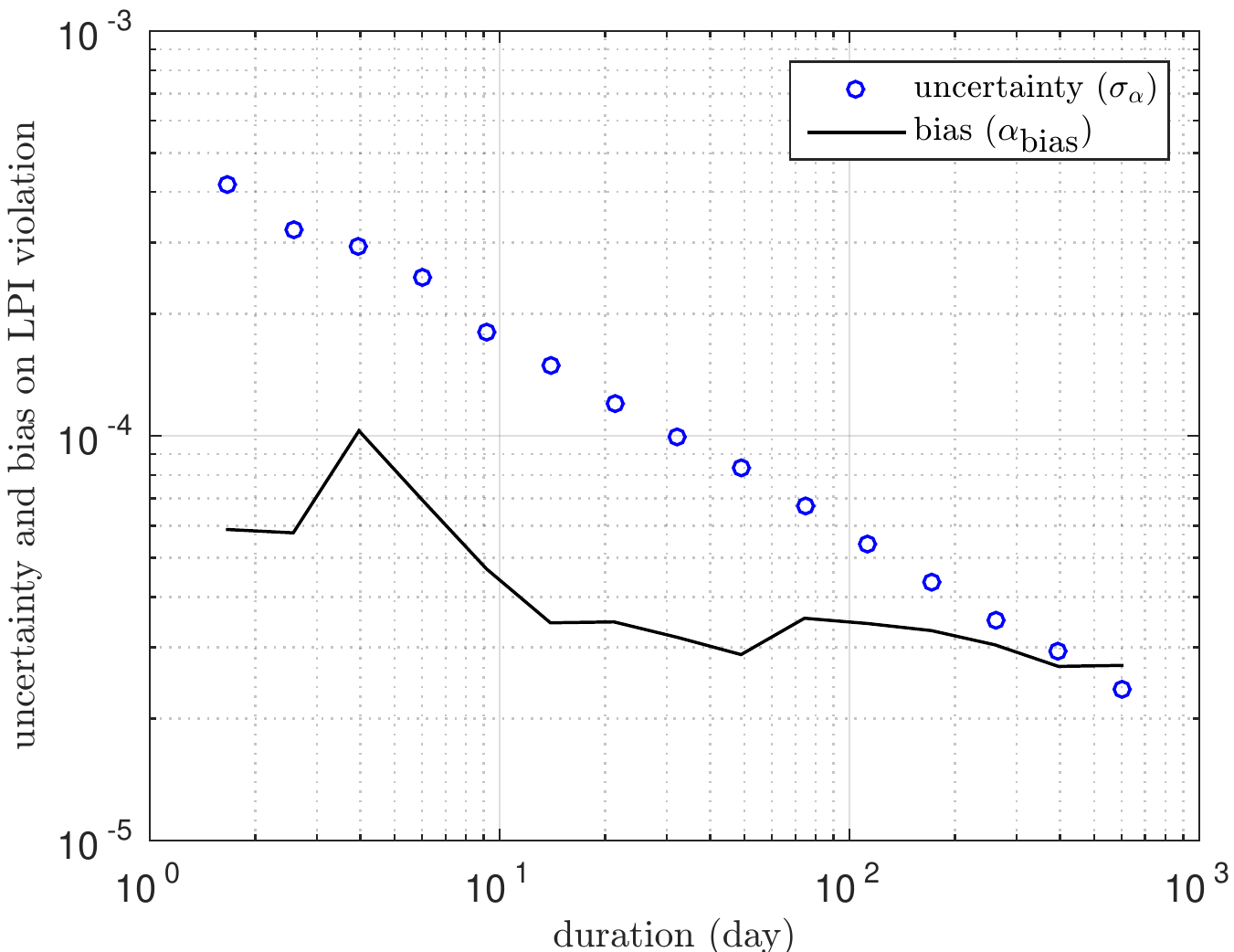}
\caption{Statistical uncertainty (blue circles) on the LPI violation and bias induced by uncorrected magnetic effect (black line) as a function of observation time, for Galileo~5\&6 final orbits.}
\label{fig:magnetic}
\end{center}
\end{figure}

To evaluate the influence of the magnetic effect on our determination of $\alpha$ we model the Earth magnetic field as a dipole with the field amplitude at the location of the satellite given by
\begin{equation} \label{equ:B}
|B|=B_0\left(\frac{R_E}{r}\right)^3\sqrt{1+3\cos^2\theta} \; ,
\end{equation}
where $B_0=0.312$~G, $R_E$ is the mean radius of the Earth and $\theta$ is the angle between the position vector of the satellite and the geomagnetic North pole. We then use the clock sensitivity coefficient to add a systematic effect $\epsilon_\textrm{sys}(t)$ to our signal, where the time dependence is given by (\ref{equ:B}) and by the satellite orbit. We determine $\alpha$ from the simulated signal $Y$ using the LSQ method (sec.~\ref{sec:stat}), such that
\begin{equation} \label{eq:Y}
Y(t) = y(t) + \epsilon_\textrm{sys}(t) + \epsilon(t) \; ,
\end{equation}
where $\epsilon$ is the random clock noise. Fig.~\ref{fig:magnetic} shows the obtained value of $\alpha$ (the bias on $\alpha$ induced by the systematic effect) as a function of observation time together with the obtained uncertainty from the LSQ-Monte Carlo method. We see that for observation times up to 450~days the magnetic effect remains below the statistical uncertainty. For longer integration times it will become limiting at $\simeq 2.7\times 10^{-5}$. This is a worst case estimation as it is of course possible to correct at least partly for the magnetic effect and/or simultaneously fit the magnetic effect and $\alpha$ (c.f. the discussion of the solar radiation pressure effect below) leading to a better result.

\begin{figure}[t] 
\begin{center}
\includegraphics[width=0.8\linewidth]{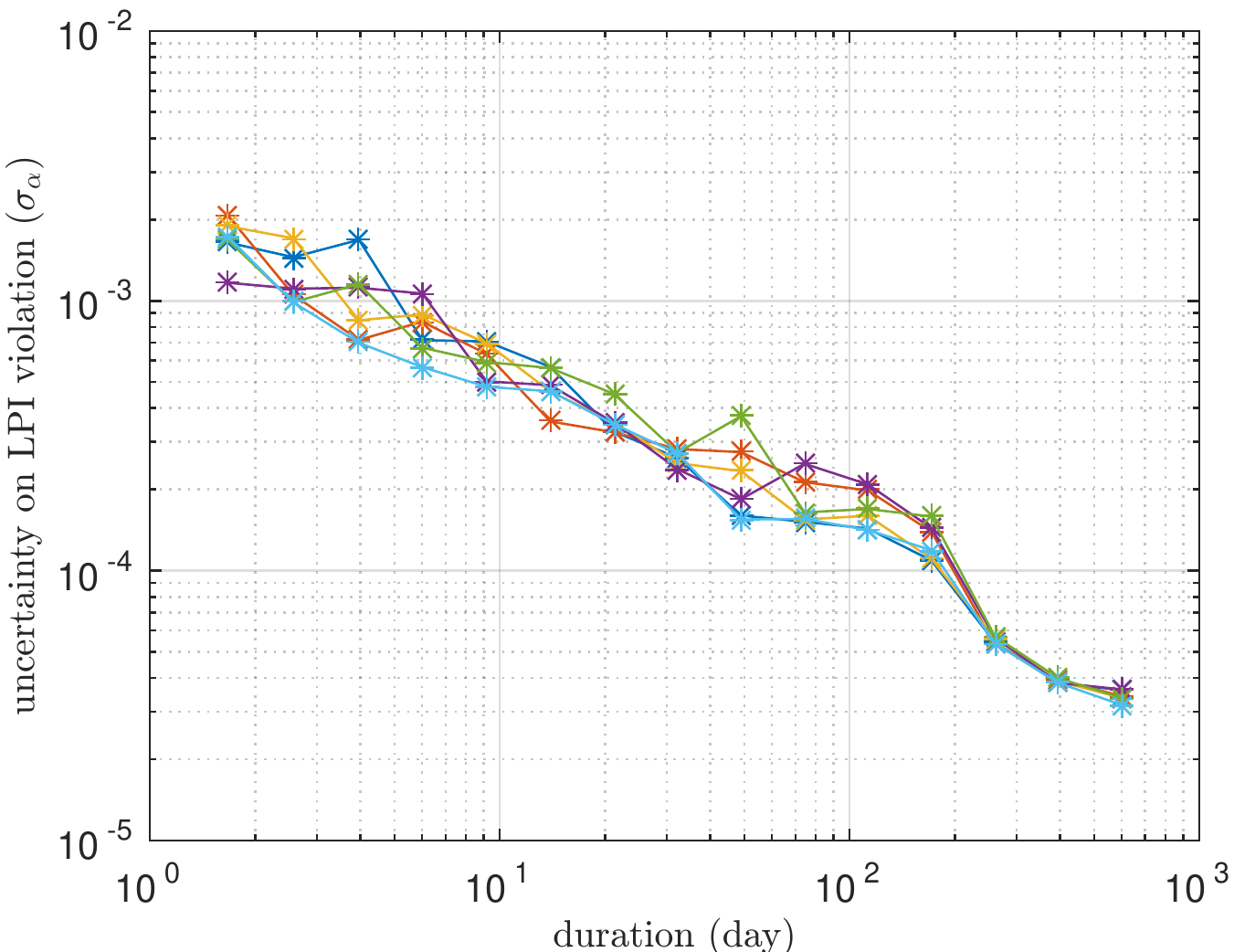}
\caption{Statistical uncertainty of the determination of the gravitational redshift in the presence of solar radiation pressure, as a function of observation time, for Galileo~5\&6 final orbits, and for six different occurrences of the simulated clock noise.}
\label{fig:SRP}
\end{center}
\end{figure}

Finally, we look at effects related to orbit modelling errors. The radial part of the resulting orbit error contributes directly to the clock estimation error. Indeed, it has been shown that the clock offset is correlated with satellite laser ranging (SLR) residuals, which can be considered as a measure of the radial orbit error~\cite{montenbruck2014}. The clock error also shows a dependency with the Sun elevation angle $\beta$, which is the angle between the satellite orbital plane and the direction of the Sun. This clearly indicates systematic effects which are linked to the direction of the Sun, as for a mismodeling of Solar Radiation Pressure (SRP)~\cite{montenbruck2014}. Indeed, the use of an enhanced SRP model for orbit determination can divide by a factor of four the effect of SRP mismodeling on the clock estimation. The resulting systematic effect on the clock estimation is evidenced as a bump at half-orbital time scale in the Modified Allan Deviation (MDEV) of Fig.~\ref{fig:clk} (left).

Consequently, we will consider here a class of systematic effects which shows a dependency with the Sun direction, as well as with the mean anomaly of the satellite. For the SRP effect in particular we use a model that reproduces the annual amplitude variation (cf. Fig.~3 of \cite{montenbruck2014}) and has a frequency of $n+\omega_\textrm{a}$, where $n$ is the mean motion of the satellite and $\omega_\textrm{a} = 2\pi / \textrm{year}$: 
\begin{equation}
\label{eq:sys}
\epsilon_\textrm{sys}(t) = A(1+B(\cos(\omega_\textrm{a}t+\phi_1)-1))\sin((n+\omega_\textrm{a})t+\phi_2) \; ,
\end{equation}
with $A,B,\phi_1,\phi_2$ constant. We simulate data with $A=5\times 10^{-14},B=0.45,\phi_1=0$ that reproduce the results of \cite{montenbruck2014}, and different values of $\phi_2$. We then use a Bayesian approach\footnote{See for example \url{http://fr.mathworks.com/help/stats/examples/bayesian-analysis-for-a-logistic-regression-model.html}} with a slice sampling method \cite{Neal2003}   assuming normal distributions and flat priors to simultaneously determine the four parameters of the SRP model (\ref{eq:sys}) and the redshift parameter $\alpha$, as well as their respective uncertainties. The result is shown in Fig.~\ref{fig:SRP} for six different occurrences of the simulated clock noise. We find that for data spanning more than $1$~year, and a ``worst case'' choice of $\phi_2$ the resulting value of $\alpha$ is unbiased with a statistical uncertainty of $\simeq 4\times 10^{-5}$, slightly increased with respect to the uncertainty when determining $\alpha$ with no SRP effect present in the observable (c.f. Fig.~\ref{fig:sensit}). After 200 days of integration time, the six occurrences of noise converge to the same values which indicates that the uncertainty is limited by the residual SRP systematic effect. In our estimates we have assumed the ``standard'' SRP corrections rather than the improved ones discussed in \cite{montenbruck2014}. The latter should allow a reduction of the effect by a factor 2-4 \cite{montenbruck2014}, so our estimates are likely to be again only an upper limit.

\section{Conclusion}

In this communication, we have shown how stable clocks placed in eccentric orbits can provide  a powerful test of the gravitational redshift. These tests are of high scientific relevance, as many alternative theories of gravitation predict a EEP violation at some level of accuracy (see for examples~\cite{damour:1994fk,damour:2010ly,damour:2012zr,minazzoli:2013fk}). In particular, we have shown that the Galileo 5 and 6 GNSS satellites can improve on the GP-A (1976) limit on the gravitational redshift test, down to an accuracy of $(3-4)\times 10^{-5}$ with at least one year of data.

Following our study, we recommend the data of Galileo satellites~5 and~6 to be made available for at least one year, in order to be able to decorrelate systematic effects from the gravitational redshift signal and to reach sufficiently low statistical uncertainties. Some data is already being collected by the IGS but ideally it should be complemented by housekeeping data such as on-board temperature, which would provide further confidence in the estimation of systematic effects.

\section*{Acknowledgments}
AH acknowledges support from ``Fonds Sp\'ecial de Recherche" through a FSR-UCL grant. PD acknowledges support from ``Action Sp\'ecifique'' GPhys/Observatoire de Paris and from the European Space Agency. The authors acknowledge L.~Prange and E.~Orliac for useful discussions, and thank the referees for their useful comments.

\section*{References}


\end{document}